\documentclass[iop]{emulateapj-rtx4}

\usepackage{graphicx}	
\usepackage{dcolumn}	
\usepackage{sistyle}		
\usepackage{times}		
\usepackage{amsmath}	
\usepackage{natbib}       	

\newcommand{\beq}{\begin{equation}}
\newcommand{\eeq}{\end{equation}}

\newcommand{\bwswitch}{0}

\shorttitle{PSR J1952+2630}
\shortauthors{Knispel et al.}

\begin{document}

\title{Arecibo PALFA Survey and Einstein@Home: Binary Pulsar Discovery by Volunteer Computing}

\author{
B.~Knispel\altaffilmark{1,2}, P.~Lazarus\altaffilmark{3}, B.~Allen\altaffilmark{1,2,4},
D.~Anderson\altaffilmark{5}, C.~Aulbert\altaffilmark{1,2}, N.~D.~R.~Bhat\altaffilmark{6},
O.~Bock\altaffilmark{1,2}, S.~Bogdanov\altaffilmark{3}, A.~Brazier\altaffilmark{7,8}, F.~Camilo\altaffilmark{9},
S.~Chatterjee\altaffilmark{7}, J.~M.~Cordes\altaffilmark{7}, F.~Crawford\altaffilmark{10},
J.~S.~Deneva\altaffilmark{11}, G.~Desvignes\altaffilmark{12}, H.~Fehrmann\altaffilmark{1,2},
P.~C.~C.~Freire\altaffilmark{13}, D.~Hammer\altaffilmark{4}, J.~W.~T.~Hessels\altaffilmark{14,15},
F.~A.~Jenet\altaffilmark{16}, V.~M.~Kaspi\altaffilmark{3}, M.~Kramer\altaffilmark{13,17},
J.~van~Leeuwen\altaffilmark{14,15}, D.~R.~Lorimer\altaffilmark{18,19}, A.~G.~Lyne\altaffilmark{17},
B.~Machenschalk\altaffilmark{1,2}, M.~A.~McLaughlin\altaffilmark{18,19}, C.~Messenger\altaffilmark{1,2,20},
D.~J.~Nice\altaffilmark{21}, M.~A.~Papa\altaffilmark{4,22}, H.~J.~Pletsch\altaffilmark{1,2},
R.~Prix\altaffilmark{1,2}, S.~M.~Ransom\altaffilmark{23}, X.~Siemens\altaffilmark{4},
I.~H.~Stairs\altaffilmark{24}, B.~W.~Stappers\altaffilmark{17}, K.~Stovall\altaffilmark{16}, and A.~Venkataraman\altaffilmark{11}}

\altaffiltext{1}{Albert-Einstein-Institut, Max-Planck-Institut f\"ur Gravitationsphysik, D-30167 Hannover, Germany}
\altaffiltext{2}{Institut f\"ur Gravitationsphysik, Leibniz Universit\"at Hannover, D-30167 Hannover, Germany}
\altaffiltext{3}{Department of Physics, McGill University, Montreal, QC H3A2T8, Canada}
\altaffiltext{4}{Physics Department, University of Wisconsin - Milwaukee, Milwaukee, WI 53211, USA} 
\altaffiltext{5}{Space Sciences Laboratory, University of California at Berkeley, Berkeley, CA 94720, USA}
\altaffiltext{6}{Swinburne University, Center for Astrophysics and Supercomputing, Hawthorn, Victoria 3122, Australia}
\altaffiltext{7}{Astronomy Department, Cornell University, Ithaca, NY 14853, USA}
\altaffiltext{8}{NAIC, Cornell University, Ithaca, NY 14853, USA}
\altaffiltext{9}{Columbia Astrophysics Laboratory, Columbia University, New York, NY 10027, USA}
\altaffiltext{10}{Department of Physics and Astronomy, Franklin and Marshall College, Lancaster, PA 17604-3003, USA}
\altaffiltext{11}{Arecibo Observatory, HC3 Box 53995, Arecibo, PR 00612, USA}
\altaffiltext{12}{Department of Astronomy and Radio Astronomy Laboratory, University of California, Berkeley, CA 94720, USA}
\altaffiltext{13}{Max-Planck-Institut f\"ur Radioastronomie, D-53121 Bonn, Germany}
\altaffiltext{14}{Netherlands Institute for Radio Astronomy (ASTRON), Postbus 2, 7990 AA Dwingeloo, The Netherlands}
\altaffiltext{15}{Astronomical Institute ``Anton Pannekoek'', University of Amsterdam, 1098 SJ Amsterdam, The Netherlands}
\altaffiltext{16}{Center for Gravitational Wave Astronomy, University Texas - Brownsville, TX 78520, USA}
\altaffiltext{17}{Jodrell Bank Centre for Astrophysics, School of Physics and Astronomy, University of Manchester, Manch., M13 9PL, UK}
\altaffiltext{18}{Department of Physics, West Virginia University, Morgantown, WV 26506, USA}
\altaffiltext{19}{Also adjunct at the NRAO (National Radio Astronomy Observatory), Green Bank, WV 24944, USA}
\altaffiltext{20}{Cardiff School of Physics and Astronomy, Cardiff University, Queens Buildings, The Parade, Cardiff, CF24 3AA, UK}
\altaffiltext{21}{Department of Physics, Lafayette College, Easton, PA 18042, USA}
\altaffiltext{22}{Albert-Einstein-Institut, Max-Planck-Institut f\"ur Gravitationsphysik, D-14476 Golm, Germany}
\altaffiltext{23}{NRAO (National Radio Astronomy Observatory), Charlottesville, VA 22903, USA}
\altaffiltext{24}{Department of Physics and Astronomy, University of British Columbia, Vancouver, BC V6T 1Z1, Canada}

\begin{abstract}
We report the discovery of the 20.7\,ms binary pulsar J1952+2630, made using the distributed 
computing project Einstein@Home in Pulsar ALFA survey observations with the Arecibo telescope.
Follow-up observations with the Arecibo telescope confirm the binary nature of the system. We obtain a 
circular orbital solution with an orbital period of \SI{9.4}{hr}, a projected orbital radius of $2.8\,\text{lt-s}$, and 
a mass function of $f=\SI{0.15}{M_\odot}$ by analysis of spin period measurements. No evidence of orbital 
eccentricity is apparent; we set a $2\sigma$ upper limit $e\lesssim\num{1.7e-3}$. The orbital parameters 
suggest a massive white dwarf companion with a minimum mass of \SI{0.95}{M_\odot}, assuming a pulsar 
mass of \SI{1.4}{M_\odot}. Most likely, this pulsar belongs to the rare class of intermediate-mass binary 
pulsars. Future timing observations will aim to determine the parameters of this system further, measure
relativistic effects, and elucidate the nature of the companion star.
\end{abstract}


\section{Introduction}\label{sec:introduction}
Pulsars in short-period orbits with neutron stars or white dwarfs are invaluable tools for 
diverse areas of science. Pulsars are precise clocks and enable very stringent tests of
Einstein's theory of general relativity (e.g., \citealp{1989ApJ...345..434T} and
\citealp{2009CQGra..26g3001K}). These binary systems also provide unique opportunities
to studying their properties as stellar-evolution endpoints (see review by \citealp{2004Sci...304..547S}). The 
detection of relativistic effects like the Shapiro delay, most easily measured for highly inclined binary 
systems with a massive companion, can reveal the orbital geometry as well as the masses of the pulsar and 
its companion \citep{2010ApJ...711..764F}. Precise mass measurements of the pulsar further our 
understanding of matter at (super)nuclear densities by providing constraints on the possible equations of 
state, e.g., \citet{2010Natur.467.1081D}. 

Here, we report the discovery and the orbital parameters of PSR J1952+2630. This pulsar
is in an almost circular 9.4\,hr orbit with a massive companion of at least \SI{0.95}{M_\odot}, 
assuming a pulsar mass of \SI{1.4}{M_\odot}. Thus, this new binary system is a good
candidate for the measurement of relativistic effects like the Shapiro delay and could have an
impact on all science areas listed above.

\section{Discovery of PSR J1952+2630}
PSR J1952+2630 was discovered in Pulsar ALFA (PALFA) survey observations taken in 2005 August with 
the 305\,m Arecibo telescope. The survey observations cover two sky regions close to the Galactic 
plane ($\left|b\right|\leq\ang{5}$). One region is in the inner Galaxy ($\ang{32} \lesssim \ell
\lesssim \ang{77}$) and is observed with \SI{268}{s} dwell time per observation; the other in the 
outer Galaxy ($\ang{168} \lesssim \ell \lesssim \ang{214}$) with \SI{134}{s} long 
observations. The observations use ALFA, a cooled seven feed-horn, dual-polarization
receiver at \SI{1.4}{GHz} \citep{2006ApJ...637..446C}. Signals are amplified, filtered, and
down-converted. Then, autocorrelation spectrometers, the Wideband Arecibo Pulsar
Processors (WAPPs; \citealp{2000ASPC..202..275D}), sum polarizations and generate spectra
over \SI{100}{MHz} of  bandwidth with 256 channels every \SI{64}{\mu s}.

Data are archived at the Center for Advanced Computing (CAC), Cornell University. They are
searched for isolated and binary pulsars in three independent pipelines: (1) a pipeline at the
CAC searching for isolated pulsars and single pulses; (2) a pipeline using the
PRESTO\footnote{\texttt{http://www.cv.nrao.edu/$\sim$sransom/presto/}} software package 
operating at several PALFA Consortium member sites, searching for isolated pulsars and 
binary pulsars with orbits longer than $\approx\SI{1}{hr}$, and single pulses; and (3) Einstein@Home, 
searching for isolated or binary pulsars with orbits longer than 11 minutes.

Einstein@Home\footnote{\texttt{http://einstein.phys.uwm.edu/}} is a distributed computing project. Its  
main goal is the detection of gravitational waves from unknown rapidly spinning  neutron stars in data 
from the LIGO and VIRGO detectors \citep{2009PhRvD..80d2003A}. Since 2009 March, about 35\% 
of Einstein@Home compute cycles have been used to search for pulsars in radio data from the 
PALFA survey. For Einstein@Home, volunteer members of the public sign up their home or office 
computers, which automatically download work units from project servers over the internet, carry out 
analyses when idle, and return results. These are automatically validated by comparison with results 
for the {\it same} work unit, produced by a different volunteer's computer. As of today, more than 
280,000 individuals have contributed; each week about 100,000 different computers download work. 
The aggregate sustained computational power of \SI{0.38}{PFlop\,s^{-1}} is comparable to that of the world's 
largest supercomputers\footnote{\texttt{http://www.top500.org/list/2010/11/100}}. 

For Einstein@Home, raw data are transferred from the CAC to the Albert Einstein Institute, Hannover, 
Germany, via high-speed internet connections. There, servers dedisperse the raw data into time series with 
628 trial dispersion measure (DM) values up to \SI{1002.4}{pc\,cm^{-3}}. For bandwidth and throughput 
reasons, the time resolution of  the raw data is reduced by a factor of two to \SI{128}{\mu s}. The 
dedispersed and downsampled time series are  downloaded by the volunteers' computers over the 
internet and coherently searched for signals from pulsars in circular orbits longer than 11 minutes. 
A detection statistic, the significance $\mathcal S = -\log_{10}\left(p\right)$, is evaluated on a grid of 
parameter space points, where $p$ is the false-alarm probability of the signal in Gaussian noise. A list 
of the 100 most  significant candidates for each dedispersed time series is returned. After completion of
all work units for a  given observation, the results are post-processed on servers in Hannover,
visually inspected  and optimized using tools from the PRESTO software package and finally uploaded to a
central database at Cornell. A more detailed account of the  Einstein@Home pipeline is available 
in \citet{J2007long}.

The 20.7-ms pulsar J1952+2630 was found with a maximum significance of $\mathcal S = 39.6$ by 
visual inspection of the Einstein@Home results from a \SI{268}{s} survey pointing with beam center at 
equatorial coordinates (J2000.0) $\alpha=19^\text{h}52^\text{m} 34.^\text{s}5$,
$\delta=\ang{+26;31;14}$. The pulsar is detected most significantly at a DM of
\SI{315.4}{pc\,cm^{-3}}. The NE2001 model of \citet{2002astro.ph..7156C} with the 
given sky position implies a  distance of $9.4^{+2.1}_{-1.4}\,\text{kpc}$. The discovery  observation 
exhibits a marginally significant, but large, barycentric period derivative $\dot P = 1.1(7)\times
10^{-9}\SI{}{s\,s^{-1}}$ over the short observation time  and indicates that PSR~J1952+2630 is in a
short-period binary system.

\section{Follow-up observations}\label{sec:discovery}
To obtain the orbital parameters of the binary system, follow-up observations with the Arecibo
telescope were carried out between 2010 July 29 and 2010 November 24. They were conducted 
mostly in coincidence with PALFA survey observations and used the central beam of the ALFA receiver with 
the Mock spectrometers\footnote{\texttt{http://www.naic.edu/$\sim$astro/mock.shtml}}. These observations 
provide a total spectral range of \SI{300}{MHz} in two overlapping bands of \SI{172}{MHz} with 512 channels 
each, centered on \SI{1.4}{GHz} at a time resolution of \SI{65.476}{\mu s}. For our analysis, we use data from 
the upper band ranging from \SI{1.364}{GHz} to \SI{1.536}{GHz}, since the lower band tends to show more 
radio frequency interference from terrestrial sources. Most follow-up observations cover an observation time 
$T_\text{obs} \approx \SI{600}{s}$, though a few are of shorter duration. On 2010 July 29 and 2010
July 30 two longer ($T_\text{obs} \approx \SI{4200}{s}$) follow-up observations with a time resolution of
\SI{142.857}{\mu s} covering the same frequency range in 2048 channels were carried out.
The epochs of the follow-up observations range from MJDs 55407 to 55525.

On 2010 August 19, gridding observations of the pulsar position were performed at \SI{2.1}{GHz} to improve 
the uncertainty of the discovery sky position. We used a square grid centered on the discovery position with 
nine observation pointings. The pulsar was found with equal signal-to-noise ratio in two of the gridding 
pointings. The improved sky position half-way between these two pointings is
$\alpha=19^\text{h}52^\text{m} 34.^\text{s}4$, $\delta=\ang{+26;30;14}$ with an uncertainty of $\approx 1'$ 
given by the telescope beam size.

Figure~\ref{fig:prof} shows the folded pulse profile of PSR J1952+2630 obtained from the 576\,s
follow-up observation on 2010 September 19. The full width at half maximum duty cycle is 6\%, 
corresponding to a width of  the pulse of $w_{50} = \SI{1.3}{ms}$. The dispersive delay across a frequency
channel is \SI{0.3}{ms} for the given DM, frequency resolution and central frequency. We calibrate the profile 
using the radiometer equation to predict the observation system's noise level. Since the sky position is 
known to $1'$ accuracy, we can only derive limits on the flux density. We obtain a system equivalent flux 
density $S_\text{sys}$ of 3.0-3.8\,Jy (depending on the precise pulsar position within 
the beam). The observation bandwidth is \SI{172}{MHz}, the observation time is \SI{576}{s}, and the folded 
profile has $N=128$ bins; then the expected off-pulse noise standard deviation used for calibration is
76-98\,$\mu$Jy. The estimated period-averaged flux density of the pulsar 
at \SI{1.4}{GHz} is $\SI{70}{\mu Jy} \leq S_{1400} \leq \SI{100}{\mu Jy}$ (depending on its position within the 
beam).
\ifcase \bwswitch
\begin{figure}
\includegraphics[width=\columnwidth]{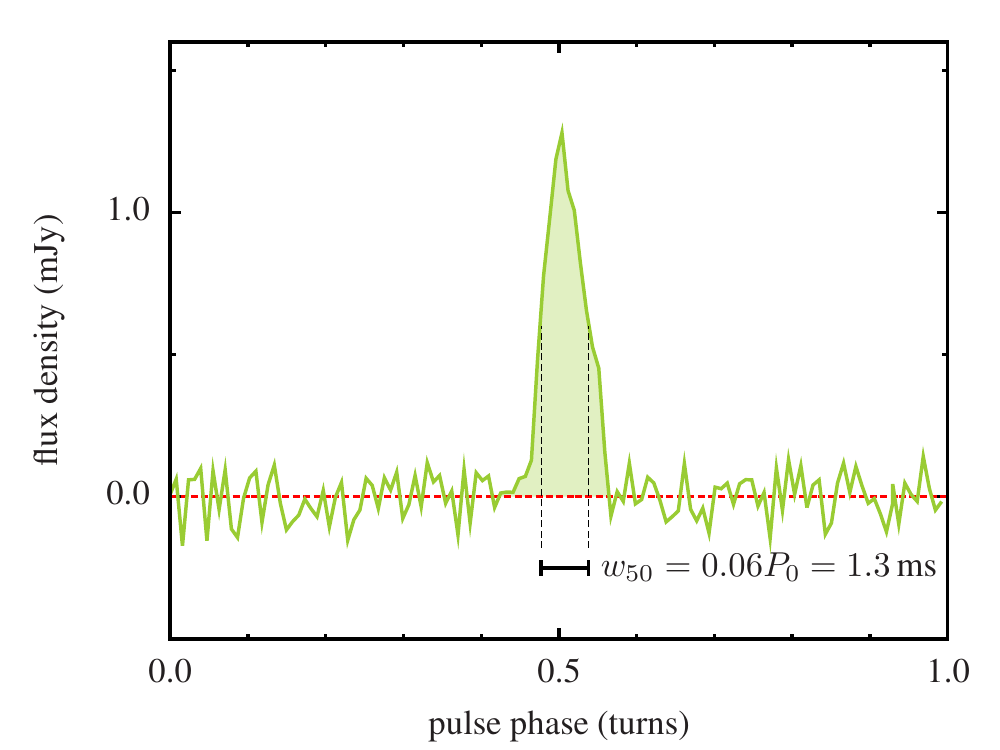}
\caption{Folded pulse profile of PSR J1952+2630 at \SI{1.4}{GHz} from the 576\,s follow-up 
observation on 2010 September 20. The resulting period-averaged flux density is $70\,\mu\text{Jy}
\leq S_{1400} \leq \SI{100}{\mu Jy}$ (depending on the precise pulsar position within the beam). The duty 
cycle is 6\% of the pulsar period $P_0$. The dispersive delay across a frequency channel is \SI{0.3}{ms}.}
\label{fig:prof}
\end{figure}
\or
\begin{figure}
\includegraphics[width=\columnwidth]{./fig1_bw.pdf}
\caption{Folded pulse profile of PSR J1952+2630 at \SI{1.4}{GHz} from the 576\,s follow-up 
observation on 2010 September 20. The resulting period-averaged flux density is $70\,\mu\text{Jy}
\leq S_{1400} \leq \SI{100}{\mu Jy}$ (depending on the precise pulsar position within the beam). The duty 
cycle is 6\% of the pulsar period $P_0$. The dispersive delay across a frequency channel is \SI{0.3}{ms}.}
\label{fig:prof}
\end{figure}
\fi

\section{Spin period-based orbital solution}\label{sec:orb}
We obtain a solution for the orbital parameters based on measurements of the barycentric spin period and 
spin period derivative. The pulsar's orbital motion in a binary system causes changes of the apparent 
barycentric spin frequency over time due to the Doppler effect. For a circular orbit, the barycentric spin period 
$P$ as a function of the orbital phase $\theta = \Omega_\text{orb}\left(t - T_\text{asc}\right)$ measured from 
the time of passage of the ascending node $T_\text{asc}$ is given by
\beq
P \left(\theta\right) = P_0\left(1 +x\Omega_\text{orb}\cos\left(\theta\right)\right).
\label{eq:period}
\eeq
Here, $\Omega_\text{orb}=2\pi/P_\text{orb}$ is the orbital angular velocity for an orbital period 
$P_\text{orb}$, $P_0$ is the intrinsic barycentric spin period of the pulsar, and $x=r \sin\left(i\right)/c$ is the 
projected orbital radius in light-seconds. The radius of the pulsar orbit is denoted by $r$, $i$ is the orbital 
inclination, and $c$ is the speed of light.

\subsection{First estimates of the orbital parameters}\label{subsec:first_est}
For each follow-up observation of length $T_\text{obs} \gtrsim \SI{600}{s}$, we obtain a pair of values, the 
barycentric spin period $P$ and the spin period derivative $\dot P$, using the PRESTO software package.
For the two longer  follow-up observations of $T_\text{obs} \approx \SI{4200}{s}$, we obtain $(P,\dot P)$ 
pairs over seven contiguous stretches of approximately \SI{600}{s} each. The acceleration is computed from
$a=c \dot P P^{-1}$, yielding in total 34 $\left(P,a\right)$ pairs. We compute the best-fitting orbital 
parameters by the method in \citet{2001MNRAS.322..885F}. We find $x =2.76(4)\,\text{lt-s}$ and
$P_\text{orb} = 9.3(1)\,\text{hr}$; numbers in parentheses are estimated $1\sigma$ errors in the last digit.

\subsection{Refining the orbital parameters}\label{subsec:refine}
The initial set of orbital parameters is extended and refined using the starting values from
Section~\ref{subsec:first_est}. We employ least-squares fitting of the full parameter set $\Omega_\text{orb}$,
$T_\text{asc}$, $x$ and $P_0$ with Equation~(\ref{eq:period}). After convergence of the fit, the parameters are 
further refined and errors are estimated by Markov Chain Monte Carlo (MCMC) sampling
\citep{2005blda.book.....G} using an independence chain Metropolis-Hastings algorithm with flat proposal 
density functions in all parameters. Table~\ref{tab:improved} shows the refined orbital parameters for  
PSR~J1952+2630  obtained in this manner.
\begin{table}
\caption{PSR J1952+2630 Parameters from a Spin-period based Analysis\label{tab:improved}}
\setlength{\extrarowheight}{1pt}
\begin{tabular*}{\columnwidth}{p{0.666\columnwidth} @{\extracolsep{\fill}}l}
\tableline
\tableline
Parameter & Value\\
\tableline
Right ascension, $\alpha$ (J2000.0)\dotfill& $19^\text{h}52.^\text{m}6$ \tablenotemark{a}\\
Declination, $\delta$ (J2000.0)\dotfill& \ang{+26;30} \tablenotemark{a}\\
Galactic longitude, $\ell$ (deg)\dotfill& 63.25  \tablenotemark{a}\\
Galactic latitude, $b$ (deg)\dotfill& $-0.37$  \tablenotemark{a}\\
Distance, $d$ (kpc) \dotfill& $9.4^{+2.1}_{-1.4}$\\
Distance from the Galactic plane, $\left|z\right|$ (kpc) \dotfill& \SI{0.06}{}\\
Dispersion measure, DM (\SI{}{pc\,cm^{-3}})\dotfill& $\SI{315.4}{}$\\
Period averaged flux density, $S_{1400}$ ($\mu$Jy) \dotfill& $ \leq100, \geq70$\\
FWHM duty cycle (pulse width, $w_{50}$ (ms))\dotfill& 6\% (\SI{1.3}{})\\
\tableline
Intrinsic barycentric spin period, $P_0$ (ms)\dotfill&$20.732368(6)\,\text{}$\\
Projected orbital radius, $x$ (lt-s)\dotfill&$2.801(3)\,\text{}$\\
Orbital period, $P_\text{orb}$ (d)\dotfill&$0.3918789(5)$\\
Time of ascending node passage, $T_\text{asc}$ (MJD)\dotfill&$55406.91066(7)$\\
Orbital eccentricity, $e$ \dotfill&$\lesssim \num{1.7e-3}$ ($2\sigma$)\\
Mass function, $f$ ($M_\odot$)\dotfill&  0.15360(1)\\
Minimum companion mass, $m_\text{c}$ ($M_\odot$)\dotfill& 0.945\\
Median companion mass, $m_\text{c, med}$ ($M_\odot$)\dotfill& 1.16\\
\tableline
\end{tabular*}
\tablenotetext{a}{The sky position was obtained from a gridding observation and has an accuracy of 1' given 
by the telescope beam size.}
\tablecomments{Thirty-four measurements of $P$ and $\dot P$ between MJDs 55407 and 55525 are used.
The numbers in parentheses show the estimated $1\sigma$ errors in the last digits.}
\end{table}

This simple model does not take into account the intrinsic spin-down of the pulsar. Fitting for the spin down 
rate is not yet possible with the data reported here, because of the degeneracy with an offset in sky position, 
which itself is not known to high enough accuracy. This effect results in a systematic error of the best-fit 
results at an unknown level. Future analysis of pulsar phase instead of spin period, will distinguish the
spin-down and orbital effects.

Figure~\ref{fig:orbphase} shows all measured spin periods $P$ as a function of the orbital phase $\theta$ in 
the upper panel and the residuals in spin period in the lower panel. No clear structure is visible in the 
residuals and thus no striking evidence of non-zero eccentricity is apparent, justifying the choice of the 
circular orbit model.
\ifcase\bwswitch
\begin{figure}
\includegraphics[width=\columnwidth]{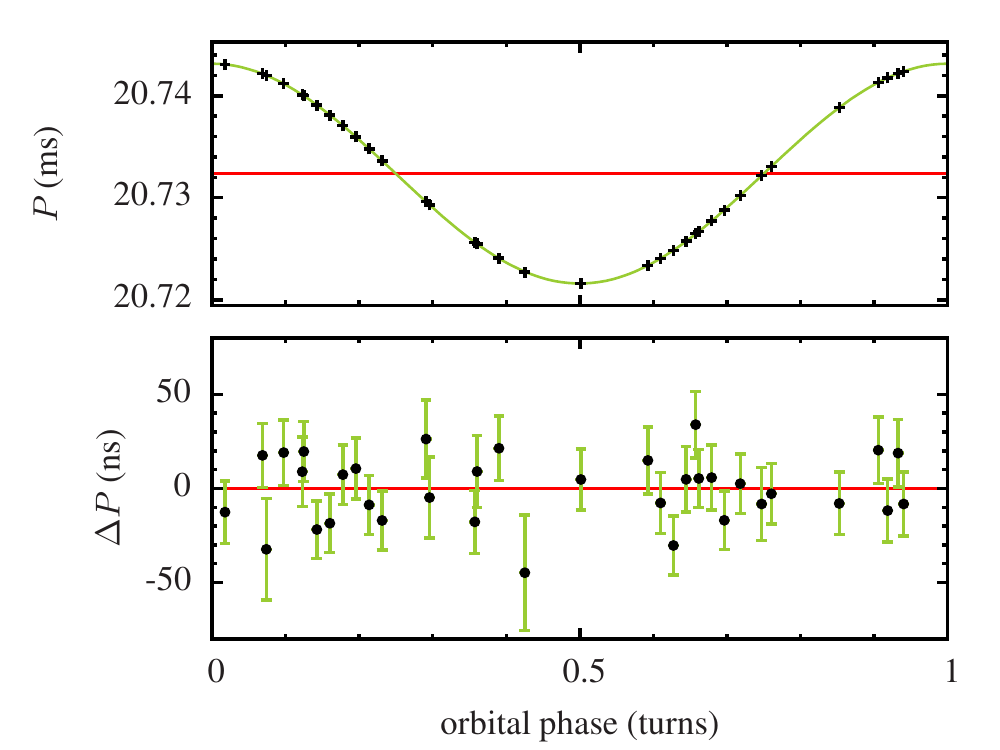}
\caption{Top: points (with errorbars, too small to see) are the measured spin periods
$P$ as a function of the orbital phase computed from the orbital model in
Table~\ref{tab:improved}. The curve is the expected model spin frequency and the horizontal red line is
at $P_0$. Bottom: the residual difference between expected and measured spin
frequency as a function of the orbital phase. No clear trend indicating non-zero eccentricity is
evident.}
\label{fig:orbphase}
\end{figure}
\or
\begin{figure}
\includegraphics[width=\columnwidth]{./fig2_bw.pdf}
\caption{Top: points (with errorbars, too small to see) are the measured spin periods
$P$ as a function of the orbital phase computed from the orbital model in
Table~\ref{tab:improved}. The curve is the expected model spin frequency and the horizontal line is
at $P_0$. Bottom: the residual difference between expected and measured spin
frequency as a function of the orbital phase. No clear trend indicating non-zero eccentricity is
evident.}
\label{fig:orbphase}
\end{figure}
\fi

\subsection{Upper limit on orbital eccentricity}\label{sec:uppecc}
To set an upper limit on the eccentricity of PSR J1952+2630's orbit, we use a spin-period model based on 
the ELL1 timing model \citep{2001MNRAS.326..274L} which is suited for orbits with small eccentricity
$e$. Then, Equation~(\ref{eq:period}) in first order in $e$ is modified to
\beq
P \left(\theta\right) = P_0\left[1 +x\Omega_\text{orb}\left( \cos\left(\theta\right) +
\varepsilon_1\sin\left(2\theta\right) +  \varepsilon_2\cos\left(2\theta\right) \right)\right],
\label{eq:ELL1}
\eeq
where the additional constants are given by $\varepsilon_1 = e\sin\left(\omega\right)$ and
$\varepsilon_2 = e\cos\left(\omega\right)$, respectively. The angle $\omega$ is the longitude of the
periastron measured with respect to the ascending node of the orbit. We now fit for the
complete set of parameters $\left(P_0, x, \Omega_\text{orb}, T_\text{asc}, \varepsilon_1,
\varepsilon_2\right)$. Following the same method as described in Section~\ref{subsec:refine}, we find the best 
fitting orbital solution including the additional eccentricity parameters. MCMC sampling of the parameters 
space is also conducted for this spin period model (Equation~(\ref{eq:ELL1})).

The parameters $\left(P_0, x, \Omega_\text{orb}, T_\text{asc}\right)$ only change slightly within the 
confidence regions obtained from the circular fit. The posterior probability distribution function (pdf) of 
the eccentricity is consistent with a circular ($e=0$) orbital model; from the posterior pdf we obtain a
$2\sigma$ upper limit on the eccentricity of $e \lesssim \num{1.7e-3}$

\section{Discussion}\label{sec:discussion}
The mass function is defined by
\beq
f = \frac{4\pi^2c^3}{G}\frac{x^3}{P_\text{orb}^2}=
\frac{\left(m_\text{c}\sin\left(i\right)\right)^3}{\left(m_\text{p} + m_\text{c}\right)^2}
\eeq
where $m_\text{p}$ and $m_\text{c}$ are the pulsar mass and the companion mass, respectively.
Inserting the orbital parameters yields $f= 0.15360(1)\,M_\odot$ for PSR~J1952+2630. Assuming a pulsar 
mass $m_\text{p} =\SI{1.4}{M_\odot}$, we obtain a minimum companion mass $m_\text{c} \geq \SI{0.945}{M_\odot}$ for 
$i=\ang{90}$. With $i=\ang{60}$ we obtain the median companion mass $m_\text{c, med} =
\SI{1.16}{M_\odot}$. For a companion mass $m_\text{c} = \SI{1.25}{M_\odot}$ (smallest measured 
neutron star mass; \citealp{2006Sci...314...97K}), we obtain an inclination angle of $i = \ang{55.1}$.

No evidence of eccentricity $e>\num{1.7e-3}$ is  found from this analysis. A common envelope phase 
with mass transfer in the system's past can explain the almost circular orbit and the short  orbital 
period. Also, the  companion's progenitor most likely was not massive enough to undergo a  supernova 
explosion; the supernova would have likely kicked the companion which almost guarantees a much higher 
orbital eccentricity. Thus, a neutron star companion is conceivable but unlikely. The companion most likely is 
a massive white dwarf. The high white dwarf companion mass, the compactness of the orbit, and the 
moderate spin period, indicate that the pulsar most likely survived a  common envelope phase
\citep{2010ApJ...711..764F}.

Low eccentricity, spin period, and high companion mass most likely place this system in the rare class of 
intermediate mass binary pulsars (IMBPs; \citealp{2001ApJ...548L.187C}). The distance of PSR J1952+2630 to 
the Galactic plane, $\left|z\right|\approx\SI{0.06}{kpc}$, is comparable to that of the other five IMBPs
\citep{2010ApJ...711..764F}; this low scale height of the known IMBP population might be due to 
observational selection effects from deep surveys along the Galactic plane \citep{2001ApJ...548L.187C}.

\citet{1992RSPTA.341...39P} derived a relation between the orbital period and eccentricity 
of low mass binary pulsars (LMBPs with companion mass $\SI{0.15}{M_\odot} \lesssim
m_\text{c}\lesssim\SI{0.4}{M_\odot}$). Figure~4 in \citet{2001ApJ...548L.187C} displays this 
relation for LMBPs and IMBPs. The LMBPs follow the theoretically predicted relation very well, 
while the IMBPs do not follow the same relation; they have higher eccentricities than LMBPs 
with the same orbital period. The slope however, seems to be very similar to the one for the LMBPs. 
This  might suggest that there exists a similar relation for this class of binary pulsars. An exact 
measurement of the orbital eccentricity of PSR~J1952+2630 from a  coherent timing solution 
will add another data point that could help to test this hypothesis at short orbital periods.

Detection of the Shapiro delay in PSR~J1952+2630 might allow precision mass estimates and strong
constraints on the orbital geometry of the binary. Figure~\ref{fig:shapiro} shows the measurable Shapiro delay 
amplitude for PSR~J1952+2630 as a function of the pulsar mass and the companion mass;  this is the 
part not absorbed by Keplerian orbital fitting (peak-to-peak amplitude of Equation~(28) in
\citealp{2010MNRAS.409..199F}).
\ifcase \bwswitch
\begin{figure}
\includegraphics[width=\columnwidth]{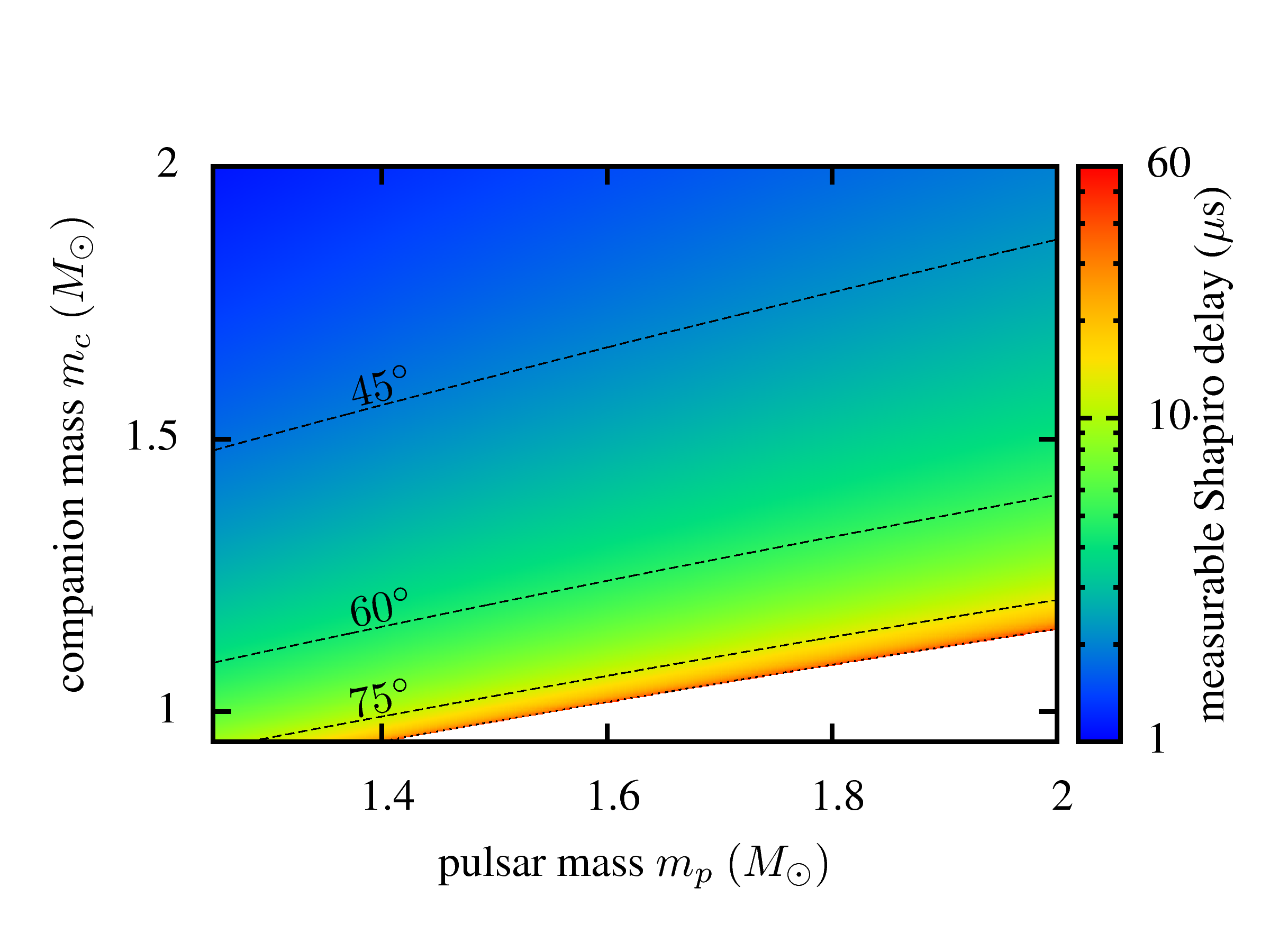}
\caption{Measurable Shapiro delay amplitude (not absorbed by Keplerian orbital fitting) for 
PSR~J1952+2630 as a function of the pulsar mass and the companion mass. The dashed lines show 
constant inclination angles of \ang{45}, \ang{60}, \ang{75}, and \ang{90} at the bottom.}
\label{fig:shapiro}
\end{figure}
\or
\begin{figure}
\includegraphics[width=\columnwidth]{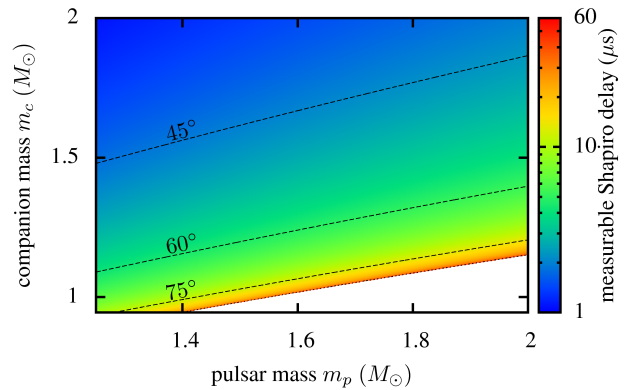}
\caption{Measurable Shapiro delay amplitude (not absorbed by Keplerian orbital fitting) for 
PSR~J1952+2630 as a function of the pulsar mass and the companion mass. The dashed lines show 
constant inclination angles of \ang{45}, \ang{60}, \ang{75}, and \ang{90} at the bottom.}
\label{fig:shapiro}
\end{figure}
\fi
Preliminary timing observations using the Mock spectrometers at the Arecibo telescope have TOA 
uncertainties of approximately \SI{20}{\mu s}. These are currently unconstraining, because no dedicated, 
deep observations at superior conjunction are available. Observations over a larger bandwidth and use of 
coherent dedispersion techniques with new instrumentation could improve TOA uncertainties further to
$\approx \SI{10}{\mu s}$. Thus, a detection of Shapiro delay requires relatively high inclination angles. If no 
Shapiro delay is detected, this will enable more stringent lower limits on the companion mass. These alone 
promise to be interesting, given the already known high minimum companion mass.

\section{Conclusions and future work}\label{sec:conclusion}
We have presented a spin-period based analysis of the orbital parameters of the newly discovered binary 
pulsar PSR J1952+2630. The pulsar has an orbital period of \SI{9.4}{hr} in an almost circular orbit with a 
projected radius of $2.8\,\text{lt-s}$. The mass function of $f=\SI{0.154}{M_\odot}$ implies a minimum 
companion mass of \SI{0.945}{M_\odot}. Most likely, the companion is a massive white dwarf, although a 
neutron star companion cannot be excluded.

The above described observations and further follow-up observations will be used to derive a coherent 
timing solution for PSR J1952+2630. This will provide a more precisely measured sky position, orbital 
parameters, and values for the orbital eccentricity and the intrinsic spin-down of the pulsar. This should
enable a detailed description of this binary system and constrain its possible formation. A precise position
would also enable searches for counterparts in X-ray, infrared, and optical wavelengths, although the
large distance makes detections challenging.

Furthermore, detection of Shapiro delay could be possible with further timing observations for high orbital 
inclinations. Given the already high minimum companion mass derived in this Letter, even a non-detection 
of the Shapiro delay could provide interesting, more stringent limits on the companion mass and its nature.

This pulsar is the second pulsar discovered by the global distributed volunteer computing
project Einstein@Home \citep{2010Sci...329.1305K}. This further demonstrates the value of
volunteer computing for discoveries in astronomy and other data-driven science.

\section*{Acknowledgements}\label{sec:acknowledgements}
We thank the Einstein@Home volunteers, who made this discovery possible. The Einstein@Home
users whose computers detected the pulsar with the highest significance are Dr.\ Vitaliy V.\  Shiryaev 
(Moscow, Russia) and Stacey Eastham (Darwen, UK).

This work was supported by CFI, CIFAR, FQRNT, MPG, NAIC,  NRAO, NSERC, NSF, NWO, and STFC. 
Arecibo is operated by the National Astronomy and Ionosphere Center under a cooperative agreement with 
the NSF. This work was supported by NSF grant AST 0807151 to Cornell University. Pulsar research at UBC 
is supported by an NSERC Discovery Grant and by the CFI. UWM and U.~C.~Berkeley acknowledge support 
by NSF grant 0555655. B.K.\ gratefully acknowledges the support of the Max Planck Society. F.C.\ 
acknowledges support from NSF grant AST-0806942. J.W.T.H.\ is a Veni Fellow of the Netherlands 
Foundation for Scientific Research (NWO). J.v.L.\ is supported by EC grant
FP7-PEOPLE-2007-4-3-IRG \#224838. D.R.L.\ and M.A.M.\ acknowledge support from a Research 
Challenge Grant from WVEPSCoR. D.J.N.\ acknowledges support by the NSF grant 0647820 to Bryn 
Mawr College.

\bibliographystyle{apj}
\bibliography{J1952+2630}

\end{document}